\documentclass[prb,twocolumn,notitlepage,longbibliography]{revtex4-2}
\usepackage{amsmath}
\usepackage{amssymb}
\usepackage{bbold} 
\usepackage{natbib}
\usepackage{lipsum}
\usepackage[unicode=true,colorlinks=true,citecolor=blue,urlcolor=blue]{hyperref}

\usepackage{bm}
\usepackage{epsfig} 
\usepackage{tikz} 
\usepackage[normalem]{ulem} 
\newcommand{\D}{\mathcal{\bm D}}

\renewcommand {\Re}{\mathop\mathrm{Re}\nolimits}

\renewcommand {\phi}{{\varphi}}
\newcommand {\rmi}{{\rm i}}

\newcommand {\e}{{\rm e}}

\begin{document}
\title{Twist tunable resonances in photonic bilayer for second harmonic generation}

\author{Egor S. Vyatkin}
\affiliation{Ioffe Institute, St. Petersburg 194021, Russia}

\author{Sergey A. Tarasenko}
\affiliation{Ioffe Institute, St. Petersburg 194021, Russia}

\email{}

\begin{abstract}
Moir\'e structure emerging in photonic bilayers stacked with a twist enables the controllable frequency selective resonant response. Here, we employ twist tunable resonances to boost second harmonic generation (SHG) at a desired frequency in twisted photonic bilayers integrated with two-dimensional nonlinear crystals. We develop an analytical theory relating the resonance frequencies and the SHG enhancement factor to the material parameters of the dielectric layers and the twist angle. The theory reveals a critical twist angle separating two distinct regimes of photonic bilayer operation: with open and closed moir\'e diffraction channels. Above the critical angle, the photon leakage from the guided mode is suppressed and the SHG enhancement factor rises by orders of magnitude. The paper offers a compact route to nonlinear conversion in moiré photonic structures efficient and tunable over a wide spectral range.

\end{abstract}
\date{\today}

\maketitle

\section{Introduction}\label{sec:intro}

Optical metasurfaces integrated with nonlinear crystals are a promising platform for efficient frequency conversion at the subwavelength scale~\cite{Autere2018,Vabishchevich2023}. By concentrating the incident electromagnetic field in high-Q evanescent modes at optical resonances, metasurfaces can strongly enhance the nonlinear response of the  crystal~\cite{Yang2015,Koshelev2019,Vyatkin2026} and tailor the emission of the generated radiation~\cite{Nookala2016,Keren-Zur2018,Marino2019}. This strategy has been demonstrated for second harmonic generation (SHG) in dielectric metasurfaces integrated with two-dimensional (2D) crystals possessing strong nonlinear susceptibilities, such as transition metal dichalcogenides~\cite{Yuan2019,Bernhardt2020,Ning2023,Khazaee2024,Ren2024}.

The design principles of 2D nonlinear structures differ from those of conventional bulk structures. In bulk crystals, efficient frequency conversion requires phase matching, so that the second harmonic waves generated by the incident wave at different points  remain in phase and add constructively~\cite{Armstrong1962}. This condition can be achieved in dispersive birefringent crystals by tuning the light propagation direction with respect to the optical axis. In 2D nonlinear structures of subwavelength thickness, the propagation length is too short for the phase accumulation to play a role. Here, efficient SHG is achieved primarily by adjusting the frequencies of optical resonances 
through the metasurface geometrical parameters.

An elegant way to control optical resonances is feasible in twisted photonic bilayers (TPB). Stacking two photonic layers and rotating them with respect to each other generate
a moir\'e pattern with a set of reciprocal vectors absent in individual layers~\cite{Tang2026}. This modifies the photonic band structure and produces high-Q guided mode resonances controlled by the interlayer distance and the twist angle~\cite{Tang2021,Lou2021,Lou2022,Salakhova2023}. 
In the linear optics regime, twist tunable photonic bilayers enable frequency filtering~\cite{Lou2022Filter}, beam steering~\cite{Lou2024,Liu2022}, and control of the spin and orbital angular momentum of light~\cite{Zhang2023,Vyatkin2025}.

Here, we employ twist tunable resonances (TTRs) to boost SHG at a desired frequency. Tuning the twist angle of a TPB plays a role similar to tuning the orientation of a bulk nonlinear crystal. We develop an analytical theory of the near field enhancement and SHG in a TPB composed of two laterally modulated dielectric layers attached to a nonlinear crystal. The theory describes the spectral position of the TTRs, the near field enhancement, and the SHG intensity in terms of the material parameters of the photonic layers and the twist angle. The analytical results and  the full-basis numerical calculations are in a good agreement. We show that there is a critical twist angle $\varphi_c$ below and above which the SHG enhancement differs by orders of magnitude. At the twist angles below $\varphi_c$, the TTR Q factor is relatively small and is limited by photon leakage from the guided modes through
the moir\'e diffraction channels.
Above the critical angle, the diffraction channels are closed, the Q factor and the SHG efficiency significantly rise and remain almost constant over the broad spectral range. Thus, twisted photonic bilayers with frequency selective and tunable guided resonances are promising for applications in nonlinear nanophotonics.

\begin{figure}
    \centering
    \includegraphics[width=\linewidth]{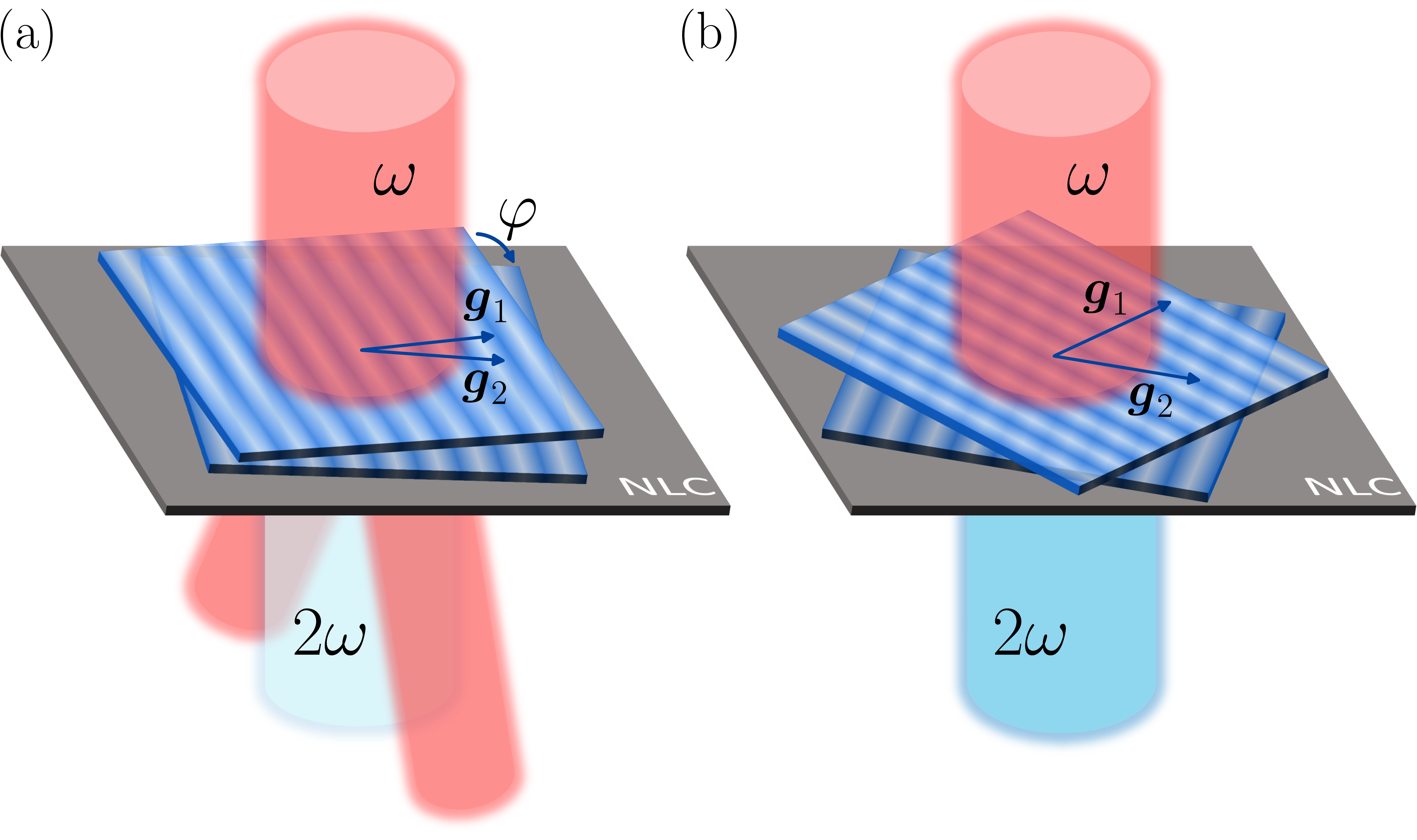}
    \caption{Twisted photonic bilayer attached to nonlinear crystal enables the resonant amplification of second harmonic generation at a desired frequency by adjusting the twist angle. Regimes of twisted photonic bilayer operation with (a) open and (b) closed moir\'e  diffraction channels.}
    \label{fig:scheme}
\end{figure}
\section{Model}

The structure we consider is a twisted photonic bilayer (TPB) integrated with a nonlinear crystal (NLC), Fig.~\ref{fig:scheme}. 

The TPB is composed of two identical thin dielectric slabs stacked with a relative twist. The dielectric permittivity of each layer $j$ ($j=1,2$) is modulated in the plane and its in-plane 2D polarizability has the form
\begin{equation}
    \alpha^{(j)}(\bm{\rho}) = \alpha_0 + \sum_{n \neq 0} \alpha_n\e^{\rmi n\bm{g}_{j} \cdot \bm{\rho}}\;,
\end{equation}
where $\alpha_0$ is the homogeneous polarizability determined by the layer thickness and the material permittivity, $\alpha_n$ are the modulation Fourier harmonics, $\bm g_j$ are the reciprocal lattice vectors, 
\[
\bm g_{1,2} = g_0 (\cos \varphi/2 ,\;\ \mp\sin \varphi/2) \,,
\]
$g_0 = |\bm g_j|$, and $\varphi$ is the twist angle, Fig.~\ref{fig:scheme}. The grating is assumed to be low contrast and centrosymmetric, i.e., $|\alpha_n|\ll \alpha_0$ and $\alpha_n=\alpha_{-n}$, respectively. Absorption in the TPB is neglected unless otherwise stated.

We consider that the reciprocal lattice vector $g_0$ is larger than the wave number of the incident light $\omega/c$ and, therefore, individual layers do not produce propagating diffracted beams at the frequency $\omega$. In the TPB geometry, however, diffracted beams can occur due to scattering by the moir\'e wave vectors $\bm g_1 \pm \bm g_2$ which can be small. 

The low-contrast TPB supports symmetric and antisymmetric (with respect to the interlayer plane) guided TE modes~\cite{Vyatkin2025}. In the limit of small interlayer distance, only the symmetric mode survives and its dispersion takes the form
\begin{equation}\label{eq:Omega_k}
    \Omega(k)=\frac{c}{2\sqrt{2}\pi(2\alpha_0)}\sqrt{\sqrt{1+[4\pi(2\alpha_0)k]^2}-1} \,,
\end{equation}
where $k$ is the in-plane wave vector. Such a structure corresponds to the metawaveguide with effective 2D polarizability 
$\alpha(\bm \rho) =  \alpha^{(1)}(\bm \rho) + \alpha^{(2)}(\bm \rho)$.

The TPB is attached to a 2D NLC where pairs of photons at the fundamental frequency $\omega$ merge into photons at the double frequency $2 \omega$. Phenomenologically, this SHG process is described by the second-order nonlinear susceptibility tensor $\hat{\chi}$, which relates the 2D polarization of the NLC at the double frequency $\bm P^{(2)}$ to the local electric field at the fundamental frequency $\bm E$ as follows
\begin{equation}
\bm P^{(2)} = \hat{\chi} \bm E \otimes \bm E \,.
\end{equation}
The polarization $\bm P^{(2)}$ acts as the source of the second harmonic emission. Note that in addition to the local term, the polarization $\bm P^{(2)}$ may contain non-local terms proportional to the field gradients. These terms contribute to the second harmonic diffracted beams which are beyond the scope of the present paper~\cite{Vyatkin2026}.

The procedure for calculating the second harmonic emission is the following. First, we solve the wave equation for the electromagnetic field at the fundamental frequency $\omega$ and determine the field distribution at the NLC attached to the TPB, see Appendix~\ref{app:A}. The near field is resonantly enhanced at the frequencies corresponding to the excitation of guided modes in the TPB. 
We consider the incident field $\bm E_{\rm in}$ polarized along 
$y \perp (\bm g_1 + \bm g_2)$, since this field is efficiently couples to the TE guided wave propagating along 
$x \parallel (\bm g_1 + \bm g_2)$.
We study the twist tunable resonances both numerically and analytically and obtain the resonance parameters in Appendices~\ref{app:B} and~\ref{app:C} for the cases of closed and open moir\'e diffraction channels, respectively. Then, using the known near field at the fundamental frequency we calculate the NLC polarization $\bm P^{(2)}$. Finally, we solve the wave equation with the source term $\bm P^{(2)}$ and calculate the intensity of second harmonic radiation emitted  by the TPB-NLC structure $I^{(2)}$, Appendix~\ref{app:D}.  

\section{Results}

\begin{figure}
    \centering
    \includegraphics[width=\linewidth]{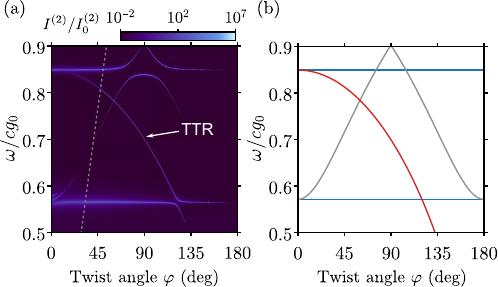}
    \caption{(a) SHG enhancement as a function of the light frequency and the twist angle. Gray dashed line shows the boundary between the areas with open and closed moir\'e diffraction channels. (b) Positions of the resonances associated with excitation of the TE waveguide modes. Red curve shows the twist tunable resonance at the frequency $\Omega(|\bm g_1 + \bm g_2|)$. The figures are calculated for the photonic bilayer parameters: $\alpha_0 g_0=0.2$, $\alpha_1 g_0 = 0.02$, and $\alpha_2 g_0 = 0.01$.}
    \label{fig:SHG-Map}
\end{figure}

Now we present and discuss the results of calculations. 

Figure~\ref{fig:SHG-Map}(a) demonstrates the color map of the SHG enhancement in the TPB-NLC structure $K = I^{(2)}/I_0^{(2)}$ as a function of the twist angle $\varphi$ and the radiation frequency $\omega$. Here, $I_0^{(2)}=8\pi \omega^2 \chi^2 E_{\rm in}^4$ is the reference intensity of second harmonic emission by the same NLC in the absence of the TPB. The enhancement exhibits resonant behavior and is observed when the incident radiation excites guided waves with the in-plane wave vectors $\bm g = n_1 \bm g_1 + n_2 \bm g_2$, where $n_1$ and $n_2$ are integers. The spectral positions of the resonances are mainly determined by the structure mean polarizability $2\alpha_0$ and the twist angle and given by $\Omega(|\bm g|)$, see Eq.~\eqref{eq:Omega_k}. Figure~\ref{fig:SHG-Map}(b) shows the positions of the resonances corresponding to $\Omega(|\bm g_1|)$ and $\Omega(2|\bm g_1|)$ (blue lines, we recall that $|\bm g_1| = |\bm g_2|$), $\Omega(|\bm g_1 + \bm g_2|)$ (red curve), and $\Omega(|2 \bm g_1 - \bm g_2|)$ and $\Omega(|2 \bm g_1 + \bm g_2|)$ (gray lines), which are visible in Fig.~\ref{fig:SHG-Map}(a). There are resonances whose spectral positions are independent of the twist angle as well as twist tunable resonances (TTRs) whose positions vary with the twist angle. The advantage of the latter is that they can be adjusted to a required frequency.

We focus on the TTR at $\Omega(|\bm g_1 + \bm g_2|) = \Omega[2 g_0 \cos (\varphi/2)]$ 
indicated by the arrow in Fig.~\ref{fig:SHG-Map}(a) and shown by the red curve in Fig.~\ref{fig:SHG-Map}(b). This TTR covers a wide spectral range and has the strongest coupling to the incident field among all TTRs in low-contrast TPBs. At the same time, its Q factor can be very high.

Near the TTR, the SHG enhancement can be presented in the form (see Appendix~\ref{app:D})
\begin{equation}\label{eq:K}
    K=\frac{I^{(2)}}{I_0^{(2)}} = \left[ \frac{c \beta |t_0(2\Omega)|}{2 \pi \Omega^2} \right]^2 \frac{\Gamma_0^2 \, \Omega^2}{[(\omega - \Omega_r)^2 + (\Gamma_0 + \Gamma)^2]^2} \,,
\end{equation}
where $\beta=-\partial\Omega/\partial(2\alpha_0)$, $t_0(\omega) = 1/(1 - 4 \pi \rmi\alpha_0 \omega/c)$ is the off-resonant amplitude transmission coefficient, 
$\Omega_r$ is the resonant frequency slightly shifted from $\Omega(|\bm g_1+\bm g_2|)$ due to dielectric grating, $\Gamma_0$ is the radiative coupling strength of the guided mode to the incident field, and $\Gamma$ is the guided mode broadening due to absorption and other possible decay channels. 

\begin{figure}
    \centering
    \includegraphics[width=\linewidth]{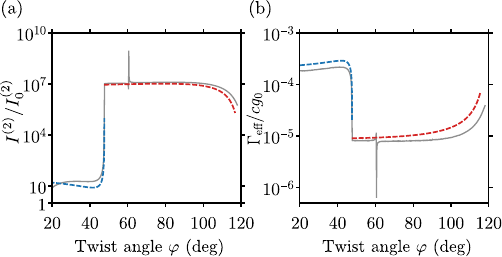}
    \caption{(a) SHG enhancement factor at twist tunable resonance and (b) half width at half maximum of the resonance as functions of the twist angle. Solid and dashed curves show the numerical and analytical results, respectively. The figures are calculated for the same parameters as Fig.~\ref{fig:SHG-Map}. }
    \label{fig:resonant-SHG}
\end{figure}

Figure~\ref{fig:resonant-SHG} shows the key parameters of the TTR: (a) the enhancement factor $K$ at $\omega = \Omega_r$ and (b) the half width at half maximum
$\Gamma_{\rm eff} = \sqrt{\sqrt{2}-1}(\Gamma_0 + \Gamma)$ as functions of the twist angle $\varphi$. Solid curves present the result of full-basis numerical calculations whereas dashed curves are plotted after analytical equations as described below. As expected, the TTR is narrow and the SHG enhancement factor can be quite large. Interestingly, the enhancement factor (as well as the TTR half width) 
differs by orders of magnitude for the twist angles $\varphi$ below and above the critical angle  
$\varphi_c = 2 \arcsin{[\Omega/(2cg_0)]}$,
$\varphi_c \approx 48 ^\circ$ in Fig.~\ref{fig:resonant-SHG}. 

The origin of this drastic difference is the moir\'e  pattern. At $\varphi < \varphi_c$, the moir\'e wave vector $|\bm g_1-\bm g_2| = 2 g_0 \sin (\varphi/2)$ is smaller than $\omega/c$, and
the diffraction channels with the in-plane wave vectors $\pm(\bm g_1-\bm g_2)$ are open, Fig.~\ref{fig:scheme}(a).
These channels provide an efficient decay pathway for guided mode excitations, thereby limiting the TTR Q factor.
(The guided and diffraction modes are coupled by the second Fourier harmonics of the permittivity modulations in the layers with the wave vectors $2 \bm g_1$ and $2 \bm g_2$ and amplitude $\alpha_2$.)
At $\varphi = \varphi_c$, the moir\'e diffraction channels get closed, Fig.~\ref{fig:scheme}(b). As a result, the SHG enhancement factor significantly rises and remains almost constant upon further increasing the twist angle.

Analytical calculations (see Appendices~\ref{app:B} and~\ref{app:C}) show that the TTR at $\Omega(|\bm g_1 + \bm g_2|)$ have the parameters
\begin{equation}\label{eq:Gamma01}
    \Gamma_0=16\pi^3 \Omega^3 \beta \Lambda^2 (\alpha_1 /c)^4 |t_0 (\Omega)|^2 \,,
\end{equation}
\begin{equation}
    \Lambda=\frac{(cg_0)^2(1-\cos \varphi)-2\Omega^2(1+4\pi \alpha_0 \varkappa)}{\Omega(1+4\pi \alpha_0 \varkappa)(c\varkappa-4\pi\alpha_0 \Omega^2/c)} \,,
\end{equation}
\begin{equation}\label{eq:Gamma}
\Gamma = 16 \pi g_0 \beta \alpha_2^2  |t_{p}(\Omega)|^2 \sqrt{\sin^2 \frac{\varphi_c}{2} - \sin^2 \frac{\varphi}{2} } \, H (\varphi_c - \varphi) \,,
\end{equation}
where $\varkappa = \sqrt{g_0^2 - (\Omega/c)^2}$, $t_p (\omega) = 1/[1 - 4 \pi \rmi \alpha_0 (\omega/c) \cos \theta_d]$ is the off-resonant transmission coefficient for a $p$-polarized wave at the moir\'e diffraction angle $\theta_d$ given by $\sin \theta_d = 2 (c g_0 /\Omega) \sin (\theta /2)$, and $H(x)$ is the Heaviside step function. The radiative coupling to the incident field $\Gamma_0$ and the diffraction-induced decay rate $\Gamma$ (at $\varphi < \varphi_c$) can be roughly estimated as $\Gamma_0 \sim \Omega (\alpha_1/\alpha_0)^4$ and $\Gamma\sim \Omega(\alpha_2/\alpha_0)^2$, respectively. 
The results of the analytical theory with $\Gamma_0$ and $\Gamma$ given by Eqs.~\eqref{eq:Gamma01} and~\eqref{eq:Gamma} are shown 
by dashed curves in Fig.~\ref{fig:resonant-SHG}. They are in good agreement with full-basis numerical calculations.   

Finally, we provide qualitative estimations for the SHG enhancement factor. At $\varphi>\varphi_c$, when the moir\'e diffraction channels are closed, the enhancement factor  is estimated as $K\sim(\Omega/\Gamma_0)^2\sim (\alpha_0/\alpha_1)^8$. In a twisted photonic bilayer with a low-contrast grating, the enhancement factor can be very high. At $\varphi<\varphi_c$, when the moir\'e diffraction channels are open, the enhancement factor is estimated as $K\sim (\Gamma_0 \, \Omega /\Gamma^2)^2 \sim (\alpha_1/\alpha_2)^8$ and determined by the ratio between the  first and second Fourier harmonics of the permittivity modulation.

\acknowledgments

This work was supported by the Russian Science Foundation (project No. 22-12-00211-$\Pi$). E.S.V. acknowledges also the support by the Foundation for Advancement of Theoretical Physics and Mathematics “BASIS”.

\appendix
\section{Field at the fundamental frequency}\label{app:A}

To calculate the near field we solve the wave equation
\begin{equation}\label{eq:maxwell}
    \operatorname{rot}\operatorname{rot} \bm E-\omega^2 \bm E=4\pi \omega^2 \alpha(\bm \rho)\delta(z) \bm E_{\parallel} 
\end{equation}
with the TPB polarizability $\alpha(\bm\rho)=\alpha^{(1)}(\bm \rho)+\alpha^{(2)}(\bm \rho)$ following Ref.~\onlinecite{Vyatkin2026}. We set $c = 1$ in Appendices, but restore the speed of light in the main text. The solution of Eq.~\eqref{eq:maxwell} can be represented  as a series
\begin{equation}
    \bm E=\bm E_{\rm in}\e^{\rmi\omega z}+\sum_{\bm g}\bm E'_{\bm g} \e^{\rmi \bm g \bm \rho+\rmi q_{g, z}|z|}\;,
\end{equation}
where $\bm E'_{g}$ denotes the Fourier harmonic of the scattered field, the sum runs over all diffraction wave vectors $\bm g=n_1 \bm g_1+n_2\bm g_2$, and $q_{g, z}=\sqrt{\omega^2-g^2}$. The in-plane components of the scattered field  
are determined by 
\begin{equation}\label{eq:E-scat}
    \bm E'_{\bm g}=\D_{\bm g}\bigg[2\alpha_0 \bm E_{\bm g}+\sum_{n\neq0}\alpha_n(\bm{E}_{\bm g-n\bm g_1}+\bm{E}_{\bm g -n \bm g_2})\bigg] \,,
\end{equation}
where $\D_{\bm g}$ is the dyadic Green function,
\begin{equation}
    \D_{\bm g} \bm E = \frac{2\pi \rmi}{q_{g, z}}
    [\omega^2 \bm E -\bm g (\bm g \cdot \bm E) ]
\end{equation}
and $\bm E_{\bm g}$ is the in-plane projection of the full field amplitude at $z=0$,
\begin{equation}\label{eq:E-f}
     \bm E_{\bm g}=\bm E'_{\bm g}+\delta_{\bm g,0}\bm E_{\rm in}\;.
 \end{equation}
Substituting Eq.~\eqref{eq:E-scat} into Eq.~\eqref{eq:E-f}, we obtain a set of linear equations, which we solve numerically and analytically in the vicinity of the TTR at $\Omega(|\bm g_1+\bm g_2|)$.

\section{Near field enhancement at closed moir\'e diffraction channels}
\label{app:B}

\begin{figure}
    \centering
    \includegraphics[width=0.89\linewidth]{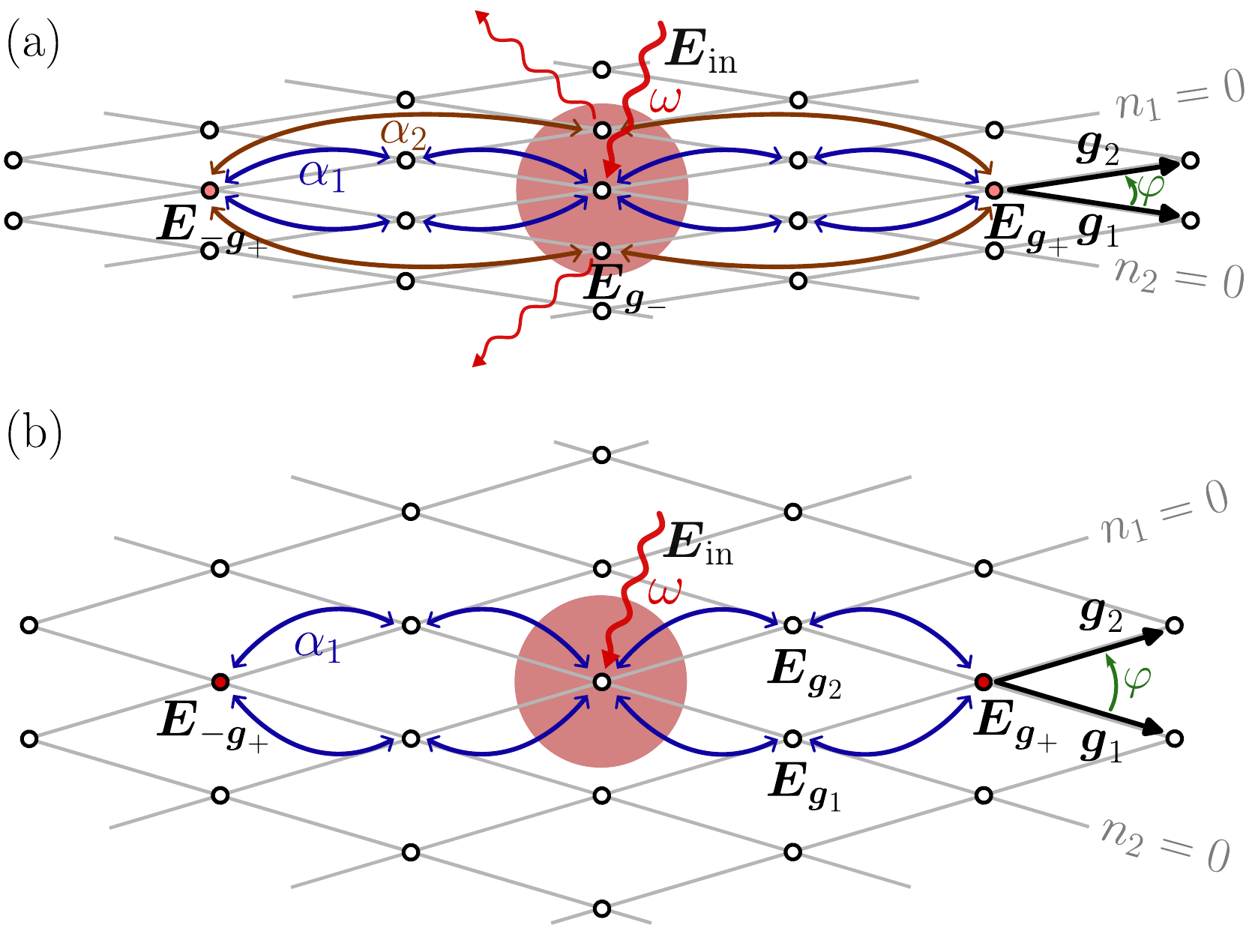}
    \caption{Fourier harmonics of the near field shown on the TPB reciprocal lattice.
    Blue and brown arcs depict the major scattering processes which determine the broadening of the TTR in the case of (a) open  and (b) closed moir\'e diffraction channels. Red circles show the light cone.}
    \label{fig:processes}
\end{figure}

Consider the resonant enhancement of the harmonics $\bm E_{\pm\bm g_+}$, where $\bm g_+=\bm g_1+\bm g_2$, at the frequency $\Omega(g_+)$ for the normal incidence of radiation. 
For the twist angles $\varphi>\varphi_c$, the radiative broadening $\Gamma_0$ is dominated by the scattering processes shown by blue arcs in Fig.~\ref{fig:processes}~(b), which are governed by the polarizability Fourier harmonic $\alpha_1$. Thus, to evaluate $\Gamma_0$, it is enough to keep only the harmonics $\alpha_0$ and $\alpha_1$ in Eq.~\eqref{eq:E-scat}. Then, equations for the field amplitudes $\bm E_0$, $\bm E_{\bm g_j}$ ($j=1,2$), and 
$\bm E_{\bm g_+}$ assume the form
\begin{eqnarray}\label{eq:E0,E1,Egp}
    \bm E_0=4\pi \rmi \omega t_0(\omega)\alpha_1[\bm E_{\bm g_1}+\bm E_{\bm g_2}]+t_0(\omega) \bm E_{\rm in} \,, \\ 
     (1-2\alpha_0\D_{\bm g_{j}})\bm E_{\bm g_{j}}=\alpha_1\D_{\bm g_{j}}(\bm E_{0}+\bm E_{\bm  g_+}) \,, \nonumber  \\
     (1-2\alpha_0\D_{\bm g_+}) \bm E_{\bm g_+}=\alpha_1 \D_{\bm g_+}(\bm E_{\bm g_1}+\bm E_{\bm g_2})\;, \nonumber
\end{eqnarray}
where we took into account that $\bm E_{\bm g}=\bm E_{-\bm g}$ and eliminated the high-order off-resonant terms $\propto \bm E_{\bm g_1 - \bm g_2}$, $\bm E_{2 \bm g_j}$, $\bm E_{2\bm g_1 + \bm g_2}$, and $\bm E_{\bm g_1 + 2\bm g_2}$. Coupling to the high-order terms leads only to a small shift $\propto \alpha_1^2$ of the resonant frequency. 

For the incident field $\bm E_{\rm in}$ polarized along $y$, $ E_{\bm g_1,y}=E_{\bm g_2,y}$, $E_{\bm g_1,x}=-E_{\bm g_2,x}$, and $\bm E_{\bm g_+} \parallel y$, as follows from Eq.~\eqref{eq:E0,E1,Egp}. Then, to the third order in $\alpha_1$, we obtain
\begin{eqnarray}\label{eq:Eg1y}
    E_{\bm g_1,y}=-\pi\alpha_1 \omega\Lambda(1-8\pi^2 \rmi \omega^2 t_0 \alpha_1^2\Lambda)(t_0 E_{\rm in}+E_{\bm g_+,y }) \,,  \nonumber \\
    (\varkappa_+ - 4 \pi \alpha_0 \omega^2) E_{\bm g_+,y} = 4\pi \alpha_1 \omega^2 
    E_{\bm g_1,y} \;, \hspace{1cm}
\end{eqnarray}
where $\varkappa_{+}=\sqrt{g_+^2-\omega^2}$ and 
\[
    \Lambda=\frac{g_0^2(1-\cos \varphi)-2\omega^2(1+4\pi \alpha_0 \varkappa)}{\omega(1+4\pi \alpha_0 \varkappa)(\varkappa-4\pi\alpha_0\omega^2)} \,.
\]
Solution of Eqs.~\eqref{eq:Eg1y} yields
\begin{equation}\label{eq:EgpSol}
     E_{\bm g_+,y}=-\frac{4\pi^2\omega^3\alpha_1^2 \Lambda t_0(\omega)  E_{\rm in}}{\varkappa_+ - 4\pi \omega^2(\alpha_0 + \delta\alpha_0)}\;,
\end{equation}
where $\alpha_0 + \delta\alpha_0$ is the renormalized coupling parameter~\cite{Vyatkin2026}. $\operatorname{Im} \delta\alpha_0 = 8\pi^3\omega^3 \alpha_1^4\Lambda^2\Re t_0(\omega)$ determines the radiative broadening $\Gamma_0$.

Near the resonance at frequency $\Omega(g_+)$, Eq.~\eqref{eq:EgpSol} can be represented in the pole form as follows 
\begin{equation}\label{eq:Egp}
     E_{\bm g_+,y}=\sqrt{\frac{\beta}{4\pi \Omega}}\frac{\sqrt{\Gamma_0}}{\omega-\Omega_r+\rmi(\Gamma_0+\Gamma)}\frac{t_0}{|t_0|} E_{\rm in} \,,
\end{equation}
where $\beta=-\partial\Omega/\partial(2\alpha_0)=8\pi^2\Omega^3\alpha_0/[1+2(4\pi\alpha_0)^2]$, 
$\Gamma_0 = 2 \beta\operatorname{Im} \delta\alpha_0$ is the radiative broadening, 
see Eq.~\eqref{eq:Gamma01}, and $\Gamma = 2\beta \operatorname{Im}\alpha_0$ is the broadening determined by absorption.
The resonant near field enhancement in the lack of  absorption is estimated as $E_{\bm g_+}/E_{\rm in} \sim \sqrt{\Omega/\Gamma_0}\sim(\alpha_0/\alpha_1)^2$.

\section{Near field enhancement at opened moir\'e diffraction channels}
\label{app:C}

At twist angles $\varphi<\varphi_c$, the excitation of the resonant guided modes with the wave vectors $\pm \bm g_{+}$ efficiently decays via the moir\'e diffraction modes with the wave vectors $\pm \bm g_{-}$ within the light cone, where $\bm g_{-} = \bm g_1 - \bm g_2$. The resonant and diffraction modes are directly coupled to each other by the polarizability Fourier harmonic $\alpha_2$. This coupling is shown by brown arcs in Fig.~\ref{fig:processes}(a). 

Taking this coupling into account, we replace in Eq.~\eqref{eq:E0,E1,Egp} the last equation for the amplitude $\bm E_{\bm g_+}$ by the set of coupled equations for the amplitudes $\bm E_{\bm g_+}$ and $\bm E_{\bm g_-}$
\begin{equation}
    (1-2\alpha_0\D_{\bm g_\pm})\bm E_{\bm g_\pm}=\D_{\bm g_\pm}
    [\alpha_1(\bm E_{\bm g_1}+\bm E_{\bm g_2}) + 2\alpha_2\bm E_{\bm g_\mp}]\;.
\end{equation}
For $E_{\bm g_+} \gg E_{\bm g_1}, E_{\bm g_2}$, the amplitude $\bm E_{\bm g_-}$ 
is given by
\begin{equation}
    E_{\bm g_-,y}=4\pi \rmi q_{g_-,z}\alpha_2 t_{p}(\omega)E_{\bm g_+,y}\;,
\end{equation}
where $t_p(\omega)=1/(1-4\pi \rmi \alpha_0 q_{g_-,z})$. 
Then, the solution of the equation set gives the amplitude $\bm E_{\bm g_+}$ in the form of Eq.~\eqref{eq:Egp} with the same $\Gamma_0$ and 
\begin{equation}
    \Gamma = 2\beta \operatorname{Im}\alpha_0 + 8 \pi q_{ g_-,z}\beta \alpha_2^2  |t_{p}(\Omega)|^2\;,
\end{equation}
see Eq.~\eqref{eq:Gamma}. The near field enhancement is estimated now as $E_{\bm g_+} / E_{\rm in} \sim (\alpha_1/\alpha_2)^2$ and is significant for $\alpha_2\ll\alpha_1$.

\section{Emission of second harmonic radiation}\label{app:D}

The polarization at the double frequency in the nonlinear crystal attached to the TPB has the form
\begin{align}
    P^{(2)}_j (\bm \rho) &= \sum_{\bm g,\bm g'}\sum_{\alpha,\beta}\chi_{jkl}E_{\bm g, k} E_{\bm g',l}  
    \e^{\rmi (\bm g+\bm g') \cdot \bm \rho}  \,.
\end{align}
The emitted field at $2 \omega$ is found from the wave equation
\begin{equation}\label{eq:2w-eq}
    \operatorname{rot}\operatorname{rot} \bm E^{(2)} - (2\omega)^2\bm E^{(2)} = 
    4\pi (2\omega)^2[\alpha(\bm \rho)\bm E^{(2)}_\parallel+\bm P^{(2)}]\delta(z)\;.
\end{equation}
with the source term $\propto \bm P^{(2)}$.

At the resonance, the near field is dominated by the harmonics $\bm E_{\pm\bm g_+} \parallel y$. Then, neglecting occasional resonant interaction at the double frequency, the amplitude of the second harmonic radiation emitted in the forward direction takes the form
\begin{equation}\label{eq:Ej_2w}
    E_{j}^{(2)} = 8\pi \rmi\omega  t_0(2\omega)\chi_{jyy}E_{\bm g_+,y}E_{-\bm g_+,y}\;.
\end{equation}
Equations~\eqref{eq:Ej_2w} and~\eqref{eq:Egp} 
yields Eq.~\eqref{eq:K} in the main text for the SHG enhancement factor.

\bibliography{SHG-twisted}

\end{document}